\documentclass[aps,twocolumn,showpacs,superscriptaddress,groupedaddress]{revtex4-1}  
\usepackage{graphicx}  
\usepackage{subfigure}
\usepackage{epstopdf}
\usepackage[pdftex]{}
\usepackage{dcolumn}   
\usepackage{bm}        
\usepackage{amssymb,amsmath,amsopn,amsfonts}
\usepackage{colordvi}
\usepackage{color}
\usepackage[none]{hyphenat}

\newcommand{\ket}[1]{\ensuremath{\left| #1 \right\rangle}}

\newcommand{\ketbra}[2]{\ensuremath{\left| #1 \rangle \langle #2\right|}}

\begin{document}

\widetext


\title{Experimentally Obtaining Maximal Coherence Via Assisted Distillation Process}

\author{Kang-Da Wu}
\thanks{These authors contributed equally to this work.}
\affiliation{Key Laboratory of Quantum Information,University of Science and Technology of China, CAS, Hefei 230026, P. R. China}
\affiliation{Synergetic Innovation Center of Quantum Information and Quantum Physics, University of Science and Technology of China, Hefei 230026, P. R. China}
\author{Zhibo Hou}
\thanks{These authors contributed equally to this work.}
\affiliation{Key Laboratory of Quantum Information,University of Science and Technology of China, CAS, Hefei 230026, P. R. China}
\affiliation{Synergetic Innovation Center of Quantum Information and Quantum Physics, University of Science and Technology of China, Hefei 230026, P. R. China}
\author{Han-Sen Zhong}
\affiliation{Key Laboratory of Quantum Information,University of Science and Technology of China, CAS, Hefei 230026, P. R. China}
\affiliation{Synergetic Innovation Center of Quantum Information and Quantum Physics, University of Science and Technology of China, Hefei 230026, P. R. China}
\author{Yuan Yuan}
\affiliation{Key Laboratory of Quantum Information,University of Science and Technology of China, CAS, Hefei 230026, P. R. China}
\affiliation{Synergetic Innovation Center of Quantum Information and Quantum Physics, University of Science and Technology of China, Hefei 230026, P. R. China}
\author{Guo-Yong Xiang}
\email{gyxiang@ustc.edu.cn}
\affiliation{Key Laboratory of Quantum Information,University of Science and Technology of China, CAS, Hefei 230026, P. R. China}
\affiliation{Synergetic Innovation Center of Quantum Information and Quantum Physics, University of Science and Technology of China, Hefei 230026, P. R. China}
\author{Chuan-Feng Li}
\affiliation{Key Laboratory of Quantum Information,University of Science and Technology of China, CAS, Hefei 230026, P. R. China}
\affiliation{Synergetic Innovation Center of Quantum Information and Quantum Physics, University of Science and Technology of China, Hefei 230026, P. R. China}
\author{Guang-Can Guo}
\affiliation{Key Laboratory of Quantum Information,University of Science and Technology of China, CAS, Hefei 230026, P. R. China}
\affiliation{Synergetic Innovation Center of Quantum Information and Quantum Physics, University of Science and Technology of China, Hefei 230026, P. R. China}

\vspace{10pt}

\date{\today}

\begin{abstract}
 Quantum coherence, which quantifies the superposition properties of a quantum state, plays an indispensable role in quantum resource theory. A recent theoretical work [Phys. Rev. Lett. \textbf{116}, 070402 (2016)] studied the manipulation of quantum coherence in bipartite or multipartite systems under the protocol Local Incoherent Operation and Classical Communication (LQICC). Here we present the first experimental realization of obtaining maximal coherence in assisted distillation protocol based on linear optical system. The results of our work show that the optimal distillable coherence rate can be reached even in one-copy scenario when the overall bipartite qubit state is pure. Moreover, the experiments for mixed states showed that distillable coherence can be increased with less demand than entanglement distillation. Our work might be helpful in the remote quantum information processing and quantum control.
\end{abstract}






\maketitle

 Quantum coherence, which exhibits the fundamental signature of superposition in quantum mechanics, has been exploited in many fields of quantum physics, such as biological systems \cite{coherencebio1,coherencebio2,coherencebio3}, transport theory \cite{coherencetrans1,coherencetrans2}, thermodynamics \cite{Thermodynamics1,Thermodynamics2,Thermodynamics3,Thermodynamics4,Thermodynamics5,Thermodynamics6,Thermodynamics7}, nanoscale physics \cite{nanoscale1} and other scientific work associated with quantum theory \cite{coherenceother1,coherenceother2,coherenceother3,coherenceother4,coherenceother5,coherenceother6,coherenceother7,coherenceother8}.
Recently, rigorous and unambiguous framework for quantifying quantum coherence has also been put forward, which has enhanced the exploitation of its operational significance in the context of quantum resource theory \cite{QuantifyingCoherence,OperationalResourceTheoryofCoherence,Yuan15intrinsic}.

Quantifying quantum coherence starts from the definition of incoherent states (free states) and incoherent operations (free operations) \cite{QuantifyingCoherence,OperationalResourceTheoryofCoherence,QRT}. A quantum state $\rho$ is  incoherent if it is diagonal in a given reference basis $\{\ket{i}\}$, i.e., $\rho=\sum_i p_i|i\rangle\langle i|$, with some probability $\{p_i\}$ \cite{MeasuringQuantumCoherencewithEntanglement,OperationalResourceTheoryofCoherence}. Incoherent operators are required to fullfill $\hat K_{n}\mathcal{I}\hat K_{n}^{\dag}\subset \mathcal{I}$ for all $n$ represented by the set of Kraus operators $\{\hat K_{n}\}$, where $\mathcal{I}$ is the set of incoherent states. Moreover, in a $d$-dimensional Hilbert space $\mathcal{H}$, the maximally coherent state is $|\Phi_d\rangle=\sqrt{1/d}\sum_i|i\rangle$, and $|\Phi\rangle:=|\Phi_2\rangle$ denotes the unit coherence resource state \cite{QuantifyingCoherence}.


 As a quantification and measure of quantum superposition in a fixed basis $\{\ket{i}\}$, various coherence measures have been proposed \cite{QuantifyingCoherence,Giro14observable,MeasuringQuantumCoherencewithEntanglement,Yuan15intrinsic}. In this paper, we choose \emph{relative entropy of coherence} \cite{QuantifyingCoherence} as the measure of this property. Relative entropy of coherence of a quantum state $\rho$ is given by $C_{r}(\rho)=S(\Delta(\rho))-S(\rho)$, where $\Delta(\rho)=\Sigma_{i}\ketbra{i}{i}\rho\ketbra{i}{i}$ is the dephasing in the reference basis. In many recent works, this kind of measure has been endowed with special significance as well as operational meaning \cite{QuantifyingCoherence,OperationalResourceTheoryofCoherence,Yuan15intrinsic}. Winter and Yang \cite{OperationalResourceTheoryofCoherence} showed that asymptotically the standard distillable coherence of a general quantum state is given by the relative entropy of quantum coherence. Likewise, Ma's group \cite{Yuan15intrinsic} has shown that the intrinsic randomness as a measure of coherence is just equal to the relative entropy of quantum coherence.


Recently, the manipulation and conversion of quantum coherence in bipartite or multipartite systems has been a hot topic \cite{Du15coherence,Ma16converting,AssistedDistillationofQuantumCoherence}. Jiajun Ma and coworkers \cite{Ma16converting} studied the interconversion of coherence and correlation between two parties under incoherent operations.  And the manipulation of coherence in a bipartite system was discussed by E. Chitambar e.t.c. \cite{AssistedDistillationofQuantumCoherence} in the task of assisted distillation of quantum coherence. The task is considered with protocols where Alice can perform arbitrary operations and Bob is restricted to only incoherent operations while classical communication between Alice and Bob is allowed, which is referred as \emph{Local Quantum-Incoherent Operations and Classical Communication} (\emph{LQICC}).  With this LQICC protocol, coherence can not be generated with respect to Bob, thus the theoretical framework forms a new kind of resource theory \cite{QRT} in which the distillable coherence on one subsystem is demanded and manipulated by the set of free operations LQICC.  And the key conclusion is also drawn out that for pure bipartite states the maximally increased distillable coherence on Bob's subsystem is equal to the von Neumann entropy of the reduced density matrix of Bob.

Quantum coherence is a highly demanded resource in many physical systems, especially in multipartite systems, however to date there is no relevant experimental implementation of assisted distillation of quantum coherence. To fill this gap, we experimentally implement a class of such LQICC protocols based on linear optic system in our laboratory, for extracting the maximal distillable coherence on one subsystem in our single-shot photonic scheme. Our results show that even in the presence of experimental imperfections and limitations, maximal increase in distillable coherence can be observed even in the single-copy scenario for pure states. Furthermore, we also reported the experimental study with a specific class of mixed states, which showed that distillable coherence can be increased with less demand than entanglement distillation.



{\bf Results}

{\bf Resouce Theory of Assisted Distillation of Quantum Coherence}. 
 In the context of quantum resource theory,  the optimal rate of generating resources (distillation rate) from a quantum state is always bounded as the manipulation and transformation are under certain restrictions \cite{QRT}. For instance, distillable coherence of a general quantum state $\rho$ under incoherent operations is equal to the relative entropy of coherence \cite{OperationalResourceTheoryofCoherence}.

The framework considered in the present task involves a bipartite system (denoted as Alice and Bob) in which coherence on Bob'side is viewed as a resource, and the set of free operations is restricted to LQICC. Under this constraint, a resource can never be generated from any \emph{quantum-incoherent} state \cite{AssistedDistillationofQuantumCoherence} (free states) which has the form of $\Sigma_i p_i\sigma^{A}_i\bigotimes\ketbra{i}{i}^B$, where $\{\ket{i}^B\}$ is the incoherent basis with respect to Bob. Our goal is to maximize distillable coherence on Bob'side with the assistance of Alice under such constraint. In order to quantify the optimal rate of preparing $\ket{\Phi}$ which can be possibly obtained, \emph{distillable coherence of collaboration} is defined as \cite{AssistedDistillationofQuantumCoherence},
\begin{equation}\label{coc}
C^{A|B}_{d}(\rho)= \sup\{R:\lim_{n\rightarrow\infty}(\inf_{\Lambda}\|\Lambda(\rho^{\bigotimes n})-\Phi^{\bigotimes\lfloor Rn\rfloor}\|)=0\},
\end{equation}
where the infimum is taken over all $\textmd{LQICC}$ operations $\Lambda$ and $\lfloor x\rfloor$ returns the maximum integer no larger than $x$.

Similarly, the bound of distillable coherence of collaboration $C^{A|B}_{d}(\rho)$ is given by the \emph{Quantum-Incoherent relative entropy} \cite{AssistedDistillationofQuantumCoherence}, i.e., ,
\begin{equation}\label{upperbound}
C^{A|B}_{d}(\rho^{AB}) \leq C^{A|B}_{r}(\rho^{AB}),
\end{equation}
which can be evaluated directly as $C^{A|B}_{r}(\rho^{AB})=S(\Delta^B(\rho^{AB}))-S(\rho^{AB})$, where $S(\cdot)$ denote the von Nuemann entropy and $\Delta^{B}$ is the dephasing in the reference incoherent basis with respect to Bob.

However, for a general bipartite resource state $\rho^{AB}$, whether the Inequality. (\ref{upperbound}) can be an equality is still unknown. In the relevant theoretical work,  E. Chitambar and coauthors have shown that at least for pure states, the equal sign is affirmative \cite{AssistedDistillationofQuantumCoherence}.

{\bf Linking One-copy Scenario to Asymptotic Settings.}  The task elaborated in the context of resource theory challenges one greatly even for a pure resource state as the optimal rate may only be obtained by working in the asymptotic settings, which involves complicated collective measurements on many copies of the shared resource states, which hinders its operational significance as well as physical implementation. Thus, linking the one-copy scenario to asymptotic settings is essential for experimentally obtaining the maximal increase in coherence.

Suppose that Alice and Bob share a pure resource state $\ket{\Psi}^{AB}$, through local measurement $M_A$ on Alice together with broadcast of the measurement outcomes, any possible pure decomposition of $\rho^{B}$ can be realized, i.e., $\rho^{B}=\Sigma_{i}p_i\ketbra{\Psi_i}{\Psi_i}$ for any set of $\{p_i\}$ and pure state $\{\ket{\Psi_i}\}$. It is obvious that Alice can help Bob get an average distillable coherence of $C_d^A(\rho^{B})=\Sigma_i p_i C_r(\ket{\Psi_i})$ in one-copy scenario (Note that $C_d(\rho)=C_r(\rho)$), which is beyond the original distillable coherence $C_d(\rho^B)$ on Bob's side as the relative entropy of coherence never increases under mixing of quantum states \cite{QuantifyingCoherence}.

For quantifying the best performance that Alice can achieve to help Bob get the maximal distillable coherence when she is restricted to just local measurement on her system and classical broadcast-processing, we use the \emph{coherence of assistance} (\textmd{COA}) \cite{AssistedDistillationofQuantumCoherence} defined as:
\begin{equation}\label{coa}
C_{a}(\rho)=\max\sum_{i}p_iS(\Delta(\Psi_i)),
\end{equation}
As shown in \cite{AssistedDistillationofQuantumCoherence}, for pure states, the coherence of collaboration $C_d^{A|B}(\ket{\Psi}^{AB})$ is equal to the regularized \textmd{COA} defined as $C^{\infty}_{a}(\rho)=\lim_{n\rightarrow\infty}\frac{1}{n}C_{a}(\rho^{\bigotimes n})=S(\Delta(\rho))$ due to higher distillation rate can be achieved through joint measurement performed on many copies of $\ket{\Psi}^{AB}$. The situation becomes more interesting as COA has additivity for a qubit state $\rho$, i.e., $C_{a}(\rho)=C^{\infty}_{a}(\rho)$ \cite{AssistedDistillationofQuantumCoherence}. Thus, obtaining the maximal increase in distillable coherence on Bob's side involves just the same local measurement applied for every copy of a purified resource state combined with classical communication. Hence, optimal LQICC only needs to be performed on every copy of the resource states instead of complex collective operations on many copies. This makes it possible to observe the maximal increase in distillable coherence in single-shot photonic experiments. Actually, the optimal LQICC to reach coherence of assistance is constructed in Appendix \ref{sec:append:Selection of measurement basis}.

\begin{figure}[!htp]
\center{\includegraphics[scale=0.27]{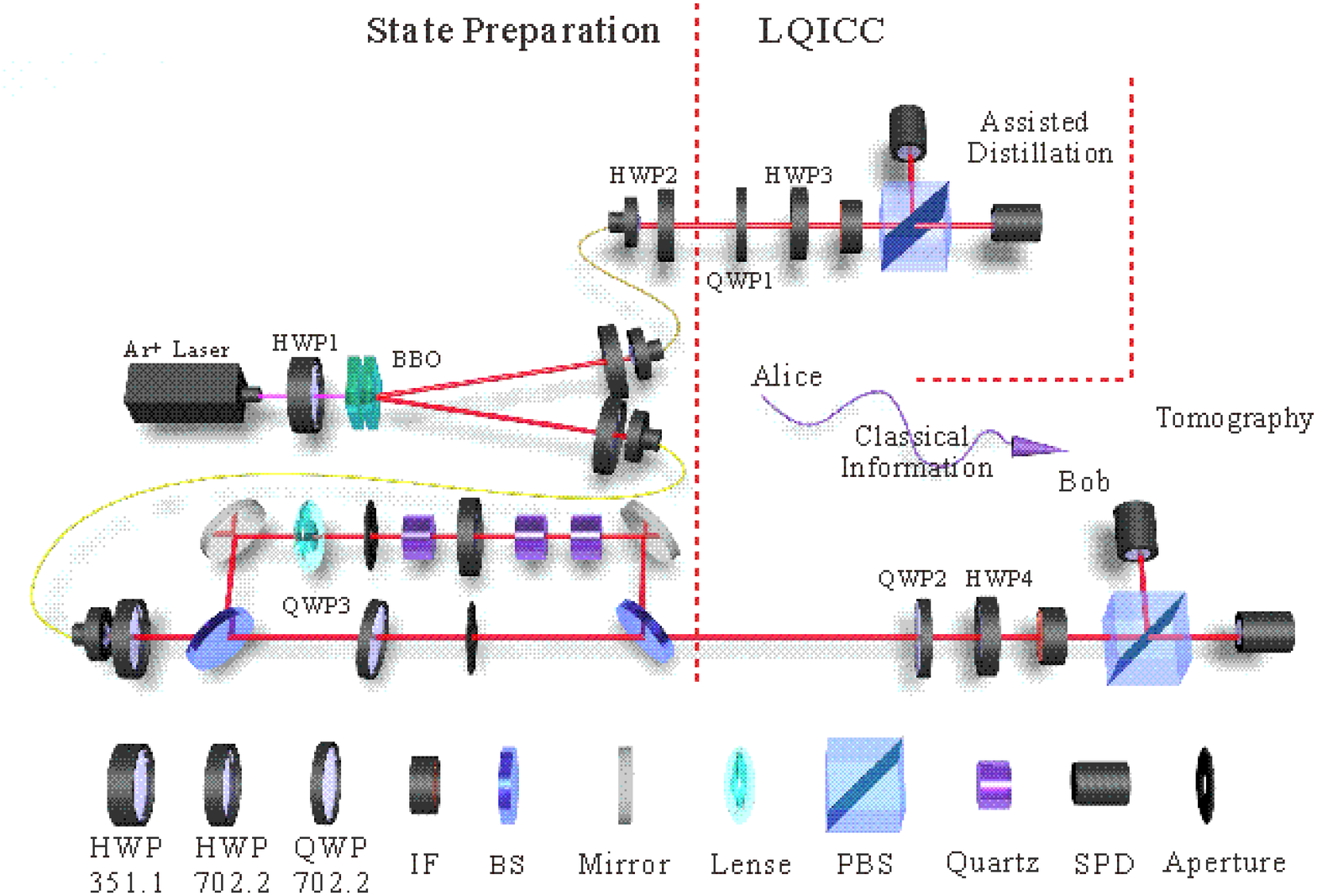}}
\caption{\label{experimental_setup}(color online) {Experimental setup. The experimental setup has two modules: state preparation, LQICC. In the state preparation module, two classes of pure states $\ket{\Psi}^{AB}=\cos2\theta\ket{HH}+\sin2\theta\ket{VV}$, $\ket{\Psi}^{AB}=\frac{1}{\sqrt{2}}(\cos2\theta\ket{HH}+\cos2\theta\ket{HV}+\sin2\theta\ket{VH}-\sin2\theta\ket{VV})$ and arbitrary Werner states in the form $\rho^{AB}=p\ketbra{\Psi^-}{\Psi^-}+(1-p)\frac{I}{4}$ can be generated. After preparation of desired two-photon resource states, the two photons are distributed to Alice and Bob. Then in the LQICC module, optimal assisted operations are performed by Alice on her photon and the measurement results are sent to Bob via classical communication channel. According to the classical message from Alice, desired corresponding incoherent operations are performed on Bob's photon. After the assisted distillation protocol, quantum state tomography is used to characterize the final state of Bob's photon to identify the final distillable coherence.  Key to components: HWP, half-wave plate; QWP, quarter-wave plate; BS, beam splitter; IF, interference filter; SPD, single photon detector; PBS, polarizing beam splitter.}}
\end{figure}
{\bf Experimental Settings and Results.}
\begin{figure}[htp]
\label{fig:first_step and second step}
\includegraphics[scale=0.49]{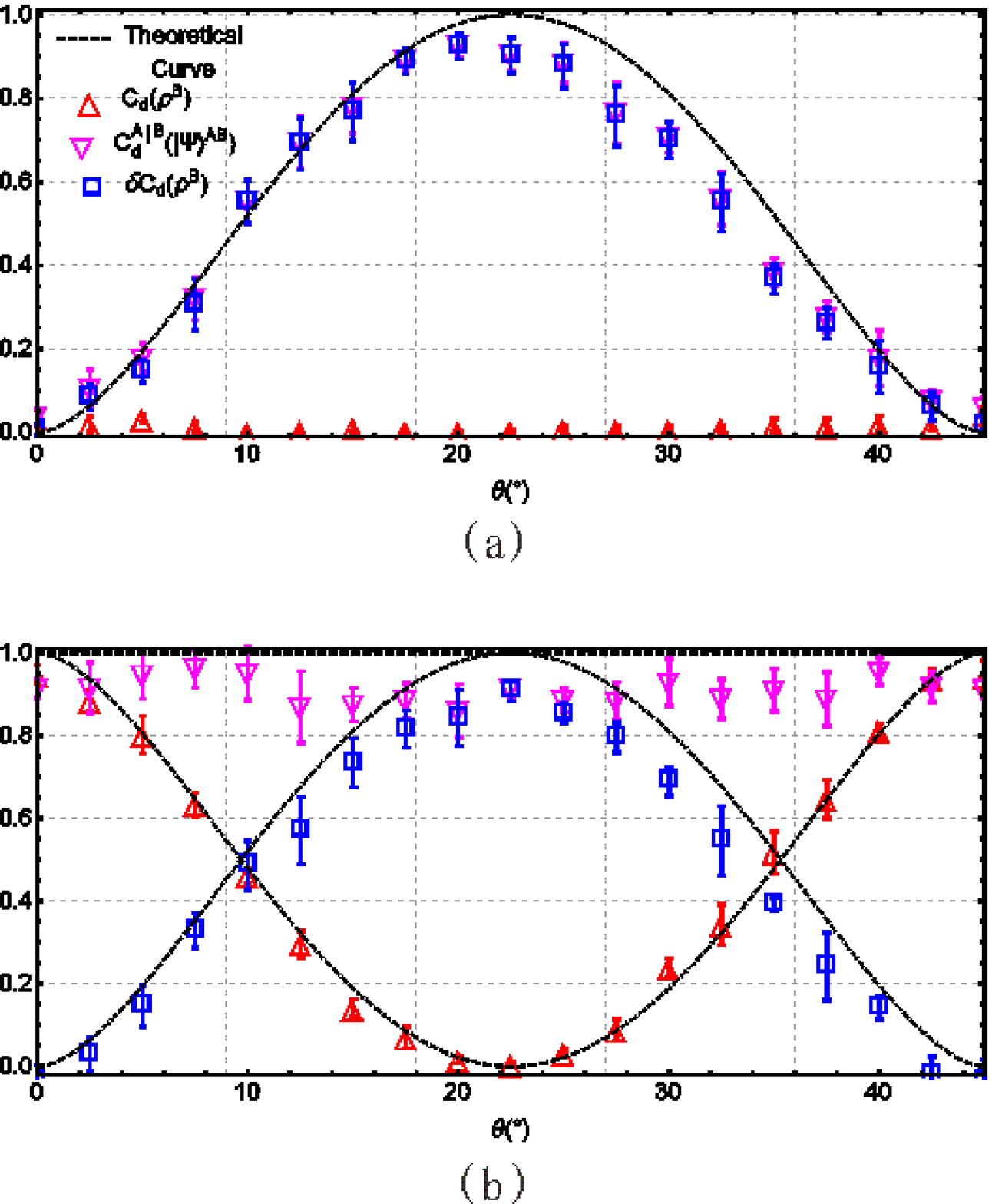}
\caption{\label{fig:all1}(Color online) Experimental results for the pure states. As shown in the figure above, $C_{d}(\rho^{B})$ (red upward-pointing triangles) represent distillable coherence of Bob without distillation, $C_{d}^{A|B}(\ket{\Psi}^{AB})$ (pink downward-pointing triangles) represent the distillable coherence of Bob after assisted distillation, $\delta C_{d}(\rho^{B})$ (blue squares) represent increase in coherence and black solid lines represent the theoretical curve. In the figure above, resource state has the form of $\ket{\Psi}^{AB}=\cos2\theta\ket{HH}+\sin2\theta\ket{VV}$ and in the figure below, $\ket{\Psi}^{AB}=\frac{1}{\sqrt{2}}(\cos2\theta\ket{HH}+\cos2\theta\ket{HV}+\sin2\theta\ket{VH}-\sin2\theta\ket{VV})$}.
\end{figure}
The whole experimental setup (details in Appendix \ref{sec:append:Experimental apparatus}) is shown in Fig. \ref{experimental_setup}. It involves two modules--state preparation module and LQICC module. In the state preparation module, pure entangled state $\ket{\Psi}^{AB}=\cos2\theta\ket{HH}+\sin2\theta\ket{VV}$ can be generated with arbitrary $\theta$ ranging from $0^\circ$ to $45^\circ$. The optical arrangement in Bob's path (see the state preparation module in Fig. \ref{experimental_setup}) is used to generate Werner states as $\rho^{AB}=p\ketbra{\Psi^-}{\Psi^-}+(1-p)\frac{I}{4}$ with an arbitrary $p$ . The LQICC module includes assisted distillation part and tomography \cite{Qi13quantum} part. In assisted distillation part, we implement the optimal local measurement on every copy of Alice's state while on Bob's side, the operation is restricted to incoherent operations. Finally Bob's state is tomographied for evaluating the final distillable coherence. For pure states, the distillable coherence of collaboration $C_d^{A|B}(\ket{\Psi}^{AB})$ can be evaluated directly by final distillable coherence of Bob. For Werner states, we replaced $C_{d}^{A|B}(\rho^{AB})$ with $C_{d}(\rho^{B}_{1})$ as we didn't know whether the distillable coherence of collaboration could be reached in single-shot experiments.

In the experimental test, we divided the procedure into two parts, respectively testing pure and mixed states in one-way assisted distillation process.

In the first part of the experiments, we generated two classes of pure states.

The first class of pure states were prepared as $\ket{\Psi}^{AB}=\cos2\theta\ket{HH}+\sin2\theta\ket{VV}$, the rotation angle $\theta$ of HWP1 was set from $0^{\circ}$ to $45^{\circ}$. Without assisted distillation, the state of Bob's photon was $\rho^B=\cos^2 2\theta\ketbra{H}{H}+\sin^2 2\theta\ketbra{V}{V}$, whose distillable coherence was evaluated as $C_d(\rho^B)=0$ theoretically. In the experiment, the state of Bob's photon was first tomographied for further calculation of distillable coherence (see red upward-pointing triangles in Fig.~\ref{fig:all1}(a)). In the assisted distillation protocol, Alice's optimal operation was to perform von Neumann measurements in the basis $\ket{\eta_\pm}^A$, which only need to be mutually unbiased with $\ket{H}$ and $\ket{V}$ for the first class of pure states (details in Appendix \ref{sec:append:Selection of measurement basis}). Here our experiment chose $\ket{\eta_\pm}^A=\ket{y_\pm}^A=\frac{1}{\sqrt{2}}(\ket{H}\pm i\ket{V})$. When Alice measured $\ket{y_+}$, the state of Bob's photon collapsed to $\cos2\theta\ket{H}-i\sin2\theta\ket{V}$ with a distillable coherence of $S(\rho^B)=-\cos^2 2\theta\log_2\cos^2 2\theta-\sin^2 2\theta\log_2\sin^2 2\theta$. When $\ket{y_-}$ was measured by Alice, the state of Bob's photon collapsed to $\cos2\theta\ket{H}+i\sin2\theta\ket{V}$ with the same amount of distillable coherence $S(\rho^B)$. After receiving Alice's measurement result (classical information) via classical communication channel, we just need to divide Bob's photons into two ensembles for future applications according to Alice's two measurement results. Since the photon in either ensemble had a distillable coherence of $S(\rho^B)$, the final distillable coherence on Bob's side after assisted distillation is $S(\rho^B)$ per photon. The distillable coherence of collaboration was evaluated as the final distillable coherence on Bob's side and denoted as $C_d^{A|B}(\ket{\Psi}^{AB})$ after its state was tomographied according to the two measurement results (see pink downward-pointing triangles in Fig.~\ref{fig:all1}(a)). The increased distillable coherence can be obtained from $\delta C_{d}(\rho^{B})=C_d^{A|B}(\ket{\Psi}^{AB})-C_d(\rho^{B})$ and equals $S(\rho^B)$ theoretically.

The second class of pure states for assisted distillation were prepared as $\ket{\Psi}^{AB}=\frac{1}{\sqrt{2}}(\cos2\theta\ket{HH}+\cos2\theta\ket{HV}+\sin2\theta\ket{VH}-\sin2\theta\ket{VV})$. Similarly, without assisted distillation the state of Bob's photon is $\rho^B=\cos^2 2\theta\ketbra{x_+}{x_+}+\sin^2 2\theta\ketbra{x_-}{x_-}$ with $\ket{x_\pm}=\frac{1}{\sqrt{2}}(\ket{H}\pm\ket{V})$, which had a distillable coherence of $C_d(\rho^{B})=\frac{1}{2}[(1+\cos 4\theta)\log_2(1+\cos 4\theta)+(1-\cos 4\theta)\log_2(1-\cos 4\theta)]$ (black dash-dotted line in Fig. \ref{fig:all1}(b). The experimental results (red upward-pointing triangles in Fig. \ref{fig:all1}(b)) were also obtained from quantum state tomography and agreed well with the theoretical predictions. In the assisted protocol, Alice's optimal operation was to perform von Neumann measurements mutually unbiased with $\cos2\theta\ket{H}+\sin2\theta\ket{V}$ and $\cos2\theta\ket{H}-\sin2\theta\ket{V}$ for the second class of pure states, and we also chose $\ket{\eta_\pm}^A=\ket{y_\pm}^A$. The state of Bob's photon collapsed to $\frac{1}{\sqrt{2}}(\cos2\theta\mp i\sin2\theta)\ket{H}+\frac{1}{\sqrt{2}}(\cos2\theta\pm i\sin2\theta)\ket{V}$ according to Alice's two measurement results $\ket{y_\pm}$ and the distillable coherence of collaboration $C_d^{A|B}(\ket{\Psi}^{AB})=1$ theoretically. We used quantum state tomography to obtain experimental results of $C_d^{A|B}(\ket{\Psi}^{AB})$ (pink downward-pointing triangles) and $\delta C_{d}(\rho^{B})$ (blue squares).

\begin{figure}[htp]
\label{fig:third_step}
\includegraphics[scale=0.49]{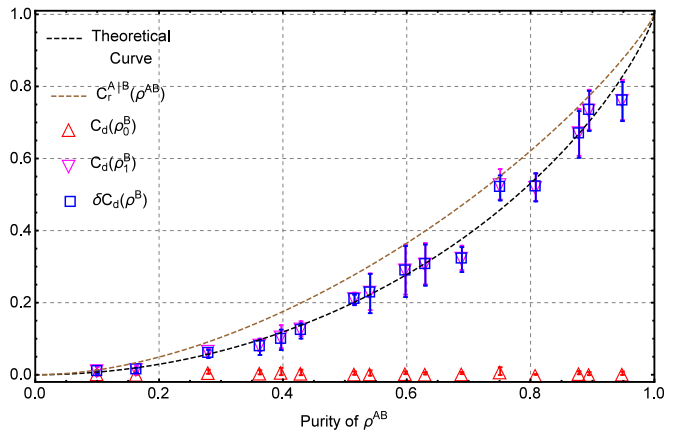}
\caption{\label{fig:all2}(Color online) Experimental results for Werner states (the colored symbols have the same meaning as shown in Fig.\ref{fig:all1}, except the brown dash-dotted line represent the theoretical calculated upper bound in inequality~(\ref{upperbound})). The resource state has the form $\rho^{AB}=p\ketbra{\Psi^-}{\Psi^-}+(1-p)\frac{I}{4}$.}
\end{figure}

In the second part of the experiments, the task of assisted distillation of coherence for mixed states is much more complicated because it is still an open problem about whether the upper bound of the Inequality~(\ref{upperbound}) can be achieved or not and what the optimal LQICC operations are. For theoretical and experimental simplicity, the optimal LQICC operation was considered for assisted distillation in one-copy instead of $n$-copy scenarios. In this part, Werner states were prepared as $\rho^{AB}=p\ketbra{\Psi^-}{\Psi^-}+(1-p)\frac{I}{4}$ with $p$ ranging from 0.05 to 0.95 by the adjustment of apertures. We denoted $\rho^B_{0}$ and $\rho^B_{1}$ as the reduced density operator for Bob before and after the assistance of Alice. Without Alice's assistance, Bob's state was a maximally mixed state, i.e., $\rho^B_{0}=\frac{I}{2}$, whose distillable coherence was $C_d(\rho^B_{0})=0$ (see the upward-pointing triangles). 
 First we found out optimal von Neumann measurements in the single-shot experiment of assisted coherence distillation, which turned out to be a maximally coherent basis $\ket{\eta_\pm}=\frac{1}{\sqrt{2}}(\ket{H}\pm e^{i\varphi}\ket{V})$ with arbitrary $\varphi$ for Werner states. For simplicity we chose $|y\pm\rangle$. After measurements along $|y\pm\rangle$ were performed on Alice's photons, Bob's state collapsed to $\rho^{B}_{1}=p\ketbra{y_\mp}{y_\mp}+(1-p)\frac{I}{2}$ according to Alice's two measurement results respectively.
Then Bob's distillable coherence after Alice's assistance was $C_d(\rho^{B}_{1})=\frac{1}{2}[(1+p)\log_2(1+p)+(1-p)\log_2(1-p)]$ theoretically (black dash-dotted line in Fig.~\ref{fig:all2}). The experimental results for $C_d(\rho^{B}_{1})$, denoted as pink downward-pointing triangles, are shown in Fig.~\ref{fig:all2}, which agreed well with the theoretical curve. The increase in distillable coherence $\delta C_d(\rho^{B})$ can be easily obtained from $\delta C_d(\rho^{B})=C_d(\rho^{B}_{1})-C_d(\rho^{B}_{0})$ (relevant experimental results denoted as blue squares in Fig.~\ref{fig:all2}). We also calculated the upper bound QI relative entropy in Eq.(\ref{upperbound}) as $C_r^{A|B}(\rho^{AB})=\frac{1}{4}[(1-p)\log_2(1-p)+(1+3p)\log_2(1+3p)-2(1+p)\log_2(1+p)]$ and the theoretical curve is shown in Fig.~\ref{fig:all2} as brown dash-dotted line.


{\bf Discussion.}


{\bf Experimental Results Analysis. }
As shown in Fig.~\ref{fig:all1}(a) and Fig.~\ref{fig:all1}(b), for the two classes of pure states, both experimental results of $C_d^{A|B}(\ket{\Psi}^{AB})$ and $\delta C_d(\rho^{B})$ are slightly smaller than the theoretical prediction around $\theta=22.5^\circ$ due to difficulty in preparing highly entangled states.

For Werner state, experimental results in Fig.\ref{fig:all2} show that the increased coherence on Bob's side in single-shot experiments agreed well with the numerical simulation and was close to the upper bound $C_{r}^{A|B}(\rho^{AB})$. However, as it is obvious that $C_d^{A|B}(\rho^{AB})\geq C_d(\rho^{B})$, whether the upper bound can be reached in the asymptotic limit is still unknown. What's interesting is that the distillable coherence on Bob's side after assisted distillation is more than zero as long as $p>0$, i.e., $C_d^{A|B}(\rho^{AB})\geq C_d(\rho^{B})>0$ iff $p>0$, which is much less demanding than the entanglement requirement with $p>1/3$ \cite{Miranowicz2004272}. Thus the distillable coherence of collaboration $C_d^{A|B}(\rho^{AB})$ is nonzero iff the state $\rho^{AB}$ is not quantum-incoherent in the reference basis. Hence, the coherence on Bob's side can still be increased with Alice's assistance even if the shared resource state $\rho^{AB}$ is not entangled.

 {\bf Conclusion }We have reported the first experimental study of assisted distillation of quantum coherence for both pure and mixed two-qubit states. In experiments for pure states, optimal assisted coherence distillation is achieved and the increased distillable coherence on Bob's side is equal to the von Neumann entropy of his quantum state. For mixed states, optimal assisted coherence distillation is considered in one-copy scenario and the maximal distillable coherence on Bob's side after assistance is close to the upper bound $C_r^{A|B}(\rho^{AB})$.

 The task of assisted distillation of quantum coherence can be applied in many scenarios. Quantum coherence is a significant property of superposition which is highly demanded in many quantum information processes such as quantum computation, one needs to distill as many copies of pure maximally coherent state as possible for better performance in these tasks. However, when considering a system where coherence is demanded is remote or unaccessible, the theoretical framework gives an optimal solution to this problem. As the results of our first experiment showed, one can extracts maximal coherence from a bipartite resource state in one-way assisted distillation process based on linear optical systems. Thus the procedure can be physically implemented in most related tasks of quantum information process with sufficiently high fidelity. Further more, the test for Werner states linked the quantum coherence to quantum correlation which is beyond entanglement, i.e., for some special form of mixed states, one can always extract distillable coherence on one subsystem as long as correlation exists, which shed light on the strong relation between quantum coherence and nonclassical correlation.



\bibliographystyle{apsrev4-1}

\bibliography{all_references_hou}

\begin{thebibliography}{33}%
\makeatletter
\providecommand \@ifxundefined [1]{%
 \@ifx{#1\undefined}
}%
\providecommand \@ifnum [1]{%
 \ifnum #1\expandafter \@firstoftwo
 \else \expandafter \@secondoftwo
 \fi
}%
\providecommand \@ifx [1]{%
 \ifx #1\expandafter \@firstoftwo
 \else \expandafter \@secondoftwo
 \fi
}%
\providecommand \natexlab [1]{#1}%
\providecommand \enquote  [1]{``#1''}%
\providecommand \bibnamefont  [1]{#1}%
\providecommand \bibfnamefont [1]{#1}%
\providecommand \citenamefont [1]{#1}%
\providecommand \href@noop [0]{\@secondoftwo}%
\providecommand \href [0]{\begingroup \@sanitize@url \@href}%
\providecommand \@href[1]{\@@startlink{#1}\@@href}%
\providecommand \@@href[1]{\endgroup#1\@@endlink}%
\providecommand \@sanitize@url [0]{\catcode `\\12\catcode `\$12\catcode
  `\&12\catcode `\#12\catcode `\^12\catcode `\_12\catcode `\%12\relax}%
\providecommand \@@startlink[1]{}%
\providecommand \@@endlink[0]{}%
\providecommand \url  [0]{\begingroup\@sanitize@url \@url }%
\providecommand \@url [1]{\endgroup\@href {#1}{\urlprefix }}%
\providecommand \urlprefix  [0]{URL }%
\providecommand \Eprint [0]{\href }%
\providecommand \doibase [0]{http://dx.doi.org/}%
\providecommand \selectlanguage [0]{\@gobble}%
\providecommand \bibinfo  [0]{\@secondoftwo}%
\providecommand \bibfield  [0]{\@secondoftwo}%
\providecommand \translation [1]{[#1]}%
\providecommand \BibitemOpen [0]{}%
\providecommand \bibitemStop [0]{}%
\providecommand \bibitemNoStop [0]{.\EOS\space}%
\providecommand \EOS [0]{\spacefactor3000\relax}%
\providecommand \BibitemShut  [1]{\csname bibitem#1\endcsname}%
\let\auto@bib@innerbib\@empty
\bibitem [{\citenamefont {Huelga}\ and\ \citenamefont
  {Plenio}(2013)}]{coherencebio1}%
  \BibitemOpen
  \bibfield  {author} {\bibinfo {author} {\bibfnamefont {S.~F.}\ \bibnamefont
  {Huelga}}\ and\ \bibinfo {author} {\bibfnamefont {M.~B.}\ \bibnamefont
  {Plenio}},\ }\href@noop {} {\bibfield  {journal} {\bibinfo  {journal}
  {Contemporary Physics}\ }\textbf {\bibinfo {volume} {54}},\ \bibinfo {pages}
  {181} (\bibinfo {year} {2013})}\BibitemShut {NoStop}%
\bibitem [{\citenamefont {Lloyd}(2011)}]{coherencebio2}%
  \BibitemOpen
  \bibfield  {author} {\bibinfo {author} {\bibfnamefont {S.}~\bibnamefont
  {Lloyd}},\ }\href {http://stacks.iop.org/1742-6596/302/i=1/a=012037}
  {\bibfield  {journal} {\bibinfo  {journal} {Journal of Physics: Conference
  Series}\ }\textbf {\bibinfo {volume} {302}},\ \bibinfo {pages} {012037}
  (\bibinfo {year} {2011})}\BibitemShut {NoStop}%
\bibitem [{\citenamefont {Plenio}\ and\ \citenamefont
  {Huelga}(2008)}]{coherencebio3}%
  \BibitemOpen
  \bibfield  {author} {\bibinfo {author} {\bibfnamefont {M.~B.}\ \bibnamefont
  {Plenio}}\ and\ \bibinfo {author} {\bibfnamefont {S.~F.}\ \bibnamefont
  {Huelga}},\ }\href {http://stacks.iop.org/1367-2630/10/i=11/a=113019}
  {\bibfield  {journal} {\bibinfo  {journal} {New Journal of Physics}\ }\textbf
  {\bibinfo {volume} {10}},\ \bibinfo {pages} {113019} (\bibinfo {year}
  {2008})}\BibitemShut {NoStop}%
\bibitem [{\citenamefont {Herranen}\ \emph {et~al.}(2009)\citenamefont
  {Herranen}, \citenamefont {Kainulainen},\ and\ \citenamefont
  {Rahkila}}]{coherencetrans1}%
  \BibitemOpen
  \bibfield  {author} {\bibinfo {author} {\bibfnamefont {M.}~\bibnamefont
  {Herranen}}, \bibinfo {author} {\bibfnamefont {K.}~\bibnamefont
  {Kainulainen}}, \ and\ \bibinfo {author} {\bibfnamefont {P.~M.}\ \bibnamefont
  {Rahkila}},\ }\href@noop {} {\bibfield  {journal} {\bibinfo  {journal}
  {Nuclear Physics A}\ }\textbf {\bibinfo {volume} {820}},\ \bibinfo {pages}
  {203c} (\bibinfo {year} {2009})}\BibitemShut {NoStop}%
\bibitem [{\citenamefont {Rebentrost}\ \emph {et~al.}(2009)\citenamefont
  {Rebentrost}, \citenamefont {Mohseni},\ and\ \citenamefont
  {Aspuru-Guzik}}]{coherencetrans2}%
  \BibitemOpen
  \bibfield  {author} {\bibinfo {author} {\bibfnamefont {P.}~\bibnamefont
  {Rebentrost}}, \bibinfo {author} {\bibfnamefont {M.}~\bibnamefont {Mohseni}},
  \ and\ \bibinfo {author} {\bibfnamefont {A.}~\bibnamefont {Aspuru-Guzik}},\
  }\href@noop {} {\bibfield  {journal} {\bibinfo  {journal} {The Journal of
  Physical Chemistry B}\ }\textbf {\bibinfo {volume} {113}},\ \bibinfo {pages}
  {9942} (\bibinfo {year} {2009})}\BibitemShut {NoStop}%
\bibitem [{\citenamefont {\AA{}berg}(2014)}]{Thermodynamics1}%
  \BibitemOpen
  \bibfield  {author} {\bibinfo {author} {\bibfnamefont {J.}~\bibnamefont
  {\AA{}berg}},\ }\href {\doibase 10.1103/PhysRevLett.113.150402} {\bibfield
  {journal} {\bibinfo  {journal} {Phys. Rev. Lett.}\ }\textbf {\bibinfo
  {volume} {113}},\ \bibinfo {pages} {150402} (\bibinfo {year}
  {2014})}\BibitemShut {NoStop}%
\bibitem [{\citenamefont {Lostaglio}\ \emph {et~al.}(2015)\citenamefont
  {Lostaglio}, \citenamefont {Korzekwa}, \citenamefont {Jennings},\ and\
  \citenamefont {Rudolph}}]{Thermodynamics2}%
  \BibitemOpen
  \bibfield  {author} {\bibinfo {author} {\bibfnamefont {M.}~\bibnamefont
  {Lostaglio}}, \bibinfo {author} {\bibfnamefont {K.}~\bibnamefont {Korzekwa}},
  \bibinfo {author} {\bibfnamefont {D.}~\bibnamefont {Jennings}}, \ and\
  \bibinfo {author} {\bibfnamefont {T.}~\bibnamefont {Rudolph}},\ }\href
  {\doibase 10.1103/PhysRevX.5.021001} {\bibfield  {journal} {\bibinfo
  {journal} {Phys. Rev. X}\ }\textbf {\bibinfo {volume} {5}},\ \bibinfo {pages}
  {021001} (\bibinfo {year} {2015})}\BibitemShut {NoStop}%
\bibitem [{\citenamefont {Korzekwa}\ \emph {et~al.}(2016)\citenamefont
  {Korzekwa}, \citenamefont {Lostaglio}, \citenamefont {Oppenheim},\ and\
  \citenamefont {Jennings}}]{Thermodynamics3}%
  \BibitemOpen
  \bibfield  {author} {\bibinfo {author} {\bibfnamefont {K.}~\bibnamefont
  {Korzekwa}}, \bibinfo {author} {\bibfnamefont {M.}~\bibnamefont {Lostaglio}},
  \bibinfo {author} {\bibfnamefont {J.}~\bibnamefont {Oppenheim}}, \ and\
  \bibinfo {author} {\bibfnamefont {D.}~\bibnamefont {Jennings}},\ }\href@noop
  {} {\bibfield  {journal} {\bibinfo  {journal} {New Journal of Physics}\
  }\textbf {\bibinfo {volume} {18}},\ \bibinfo {pages} {023045} (\bibinfo
  {year} {2016})}\BibitemShut {NoStop}%
\bibitem [{\citenamefont {\ifmmode \acute{C}\else
  \'{C}\fi{}wikli\ifmmode~\acute{n}\else \'{n}\fi{}ski}\ \emph
  {et~al.}(2015)\citenamefont {\ifmmode \acute{C}\else
  \'{C}\fi{}wikli\ifmmode~\acute{n}\else \'{n}\fi{}ski}, \citenamefont
  {Studzi\ifmmode~\acute{n}\else \'{n}\fi{}ski}, \citenamefont {Horodecki},\
  and\ \citenamefont {Oppenheim}}]{Thermodynamics4}%
  \BibitemOpen
  \bibfield  {author} {\bibinfo {author} {\bibfnamefont {P.}~\bibnamefont
  {\ifmmode \acute{C}\else \'{C}\fi{}wikli\ifmmode~\acute{n}\else
  \'{n}\fi{}ski}}, \bibinfo {author} {\bibfnamefont {M.}~\bibnamefont
  {Studzi\ifmmode~\acute{n}\else \'{n}\fi{}ski}}, \bibinfo {author}
  {\bibfnamefont {M.}~\bibnamefont {Horodecki}}, \ and\ \bibinfo {author}
  {\bibfnamefont {J.}~\bibnamefont {Oppenheim}},\ }\href {\doibase
  10.1103/PhysRevLett.115.210403} {\bibfield  {journal} {\bibinfo  {journal}
  {Phys. Rev. Lett.}\ }\textbf {\bibinfo {volume} {115}},\ \bibinfo {pages}
  {210403} (\bibinfo {year} {2015})}\BibitemShut {NoStop}%
\bibitem [{\citenamefont {Narasimhachar}\ and\ \citenamefont
  {Gour}(2015)}]{Thermodynamics5}%
  \BibitemOpen
  \bibfield  {author} {\bibinfo {author} {\bibfnamefont {V.}~\bibnamefont
  {Narasimhachar}}\ and\ \bibinfo {author} {\bibfnamefont {G.}~\bibnamefont
  {Gour}},\ }\href@noop {} {\bibfield  {journal} {\bibinfo  {journal} {Nature
  communications}\ }\textbf {\bibinfo {volume} {6}} (\bibinfo {year}
  {2015})}\BibitemShut {NoStop}%
\bibitem [{\citenamefont {Rodr{\'\i}guez-Rosario}\ \emph
  {et~al.}(2013)\citenamefont {Rodr{\'\i}guez-Rosario}, \citenamefont
  {Frauenheim},\ and\ \citenamefont {Aspuru-Guzik}}]{Thermodynamics6}%
  \BibitemOpen
  \bibfield  {author} {\bibinfo {author} {\bibfnamefont {C.~A.}\ \bibnamefont
  {Rodr{\'\i}guez-Rosario}}, \bibinfo {author} {\bibfnamefont {T.}~\bibnamefont
  {Frauenheim}}, \ and\ \bibinfo {author} {\bibfnamefont {A.}~\bibnamefont
  {Aspuru-Guzik}},\ }\href@noop {} {\bibfield  {journal} {\bibinfo  {journal}
  {arXiv:1308.1245}\ } (\bibinfo {year} {2013})}\BibitemShut {NoStop}%
\bibitem [{\citenamefont {Gardas}\ and\ \citenamefont
  {Deffner}(2015)}]{Thermodynamics7}%
  \BibitemOpen
  \bibfield  {author} {\bibinfo {author} {\bibfnamefont {B.}~\bibnamefont
  {Gardas}}\ and\ \bibinfo {author} {\bibfnamefont {S.}~\bibnamefont
  {Deffner}},\ }\href {\doibase 10.1103/PhysRevE.92.042126} {\bibfield
  {journal} {\bibinfo  {journal} {Phys. Rev. E}\ }\textbf {\bibinfo {volume}
  {92}},\ \bibinfo {pages} {042126} (\bibinfo {year} {2015})}\BibitemShut
  {NoStop}%
\bibitem [{\citenamefont {Karlstr\"om}\ \emph {et~al.}(2011)\citenamefont
  {Karlstr\"om}, \citenamefont {Linke}, \citenamefont {Karlstr\"om},\ and\
  \citenamefont {Wacker}}]{nanoscale1}%
  \BibitemOpen
  \bibfield  {author} {\bibinfo {author} {\bibfnamefont {O.}~\bibnamefont
  {Karlstr\"om}}, \bibinfo {author} {\bibfnamefont {H.}~\bibnamefont {Linke}},
  \bibinfo {author} {\bibfnamefont {G.}~\bibnamefont {Karlstr\"om}}, \ and\
  \bibinfo {author} {\bibfnamefont {A.}~\bibnamefont {Wacker}},\ }\href
  {\doibase 10.1103/PhysRevB.84.113415} {\bibfield  {journal} {\bibinfo
  {journal} {Phys. Rev. B}\ }\textbf {\bibinfo {volume} {84}},\ \bibinfo
  {pages} {113415} (\bibinfo {year} {2011})}\BibitemShut {NoStop}%
\bibitem [{\citenamefont {Chen}\ \emph {et~al.}(2015)\citenamefont {Chen},
  \citenamefont {Cui},\ and\ \citenamefont {Fan}}]{coherenceother1}%
  \BibitemOpen
  \bibfield  {author} {\bibinfo {author} {\bibfnamefont {J.-J.}\ \bibnamefont
  {Chen}}, \bibinfo {author} {\bibfnamefont {J.}~\bibnamefont {Cui}}, \ and\
  \bibinfo {author} {\bibfnamefont {H.}~\bibnamefont {Fan}},\ }\href@noop {}
  {\bibfield  {journal} {\bibinfo  {journal} {arXiv:1509.03576}\ } (\bibinfo
  {year} {2015})}\BibitemShut {NoStop}%
\bibitem [{\citenamefont {Cheng}\ and\ \citenamefont
  {Hall}(2015)}]{coherenceother2}%
  \BibitemOpen
  \bibfield  {author} {\bibinfo {author} {\bibfnamefont {S.}~\bibnamefont
  {Cheng}}\ and\ \bibinfo {author} {\bibfnamefont {M.~J.~W.}\ \bibnamefont
  {Hall}},\ }\href {\doibase 10.1103/PhysRevA.92.042101} {\bibfield  {journal}
  {\bibinfo  {journal} {Phys. Rev. A}\ }\textbf {\bibinfo {volume} {92}},\
  \bibinfo {pages} {042101} (\bibinfo {year} {2015})}\BibitemShut {NoStop}%
\bibitem [{\citenamefont {Hu}\ and\ \citenamefont
  {Fan}(2015)}]{coherenceother3}%
  \BibitemOpen
  \bibfield  {author} {\bibinfo {author} {\bibfnamefont {X.}~\bibnamefont
  {Hu}}\ and\ \bibinfo {author} {\bibfnamefont {H.}~\bibnamefont {Fan}},\
  }\href@noop {} {\bibfield  {journal} {\bibinfo  {journal} {arXiv:1508.01978}\
  } (\bibinfo {year} {2015})}\BibitemShut {NoStop}%
\bibitem [{\citenamefont {Marvian}\ \emph {et~al.}(2016)\citenamefont
  {Marvian}, \citenamefont {Spekkens},\ and\ \citenamefont
  {Zanardi}}]{coherenceother4}%
  \BibitemOpen
  \bibfield  {author} {\bibinfo {author} {\bibfnamefont {I.}~\bibnamefont
  {Marvian}}, \bibinfo {author} {\bibfnamefont {R.~W.}\ \bibnamefont
  {Spekkens}}, \ and\ \bibinfo {author} {\bibfnamefont {P.}~\bibnamefont
  {Zanardi}},\ }\href {\doibase 10.1103/PhysRevA.93.052331} {\bibfield
  {journal} {\bibinfo  {journal} {Phys. Rev. A}\ }\textbf {\bibinfo {volume}
  {93}},\ \bibinfo {pages} {052331} (\bibinfo {year} {2016})}\BibitemShut
  {NoStop}%
\bibitem [{\citenamefont {Bai}\ and\ \citenamefont
  {Du}(2015)}]{coherenceother5}%
  \BibitemOpen
  \bibfield  {author} {\bibinfo {author} {\bibfnamefont {Z.}~\bibnamefont
  {Bai}}\ and\ \bibinfo {author} {\bibfnamefont {S.}~\bibnamefont {Du}},\
  }\href@noop {} {\bibfield  {journal} {\bibinfo  {journal} {arXiv:1503.07103}\
  } (\bibinfo {year} {2015})}\BibitemShut {NoStop}%
\bibitem [{\citenamefont {Liu}\ \emph {et~al.}(2016)\citenamefont {Liu},
  \citenamefont {Tian}, \citenamefont {Wang},\ and\ \citenamefont
  {Jing}}]{coherenceother6}%
  \BibitemOpen
  \bibfield  {author} {\bibinfo {author} {\bibfnamefont {X.}~\bibnamefont
  {Liu}}, \bibinfo {author} {\bibfnamefont {Z.}~\bibnamefont {Tian}}, \bibinfo
  {author} {\bibfnamefont {J.}~\bibnamefont {Wang}}, \ and\ \bibinfo {author}
  {\bibfnamefont {J.}~\bibnamefont {Jing}},\ }\href@noop {} {\bibfield
  {journal} {\bibinfo  {journal} {Annals of Physics}\ }\textbf {\bibinfo
  {volume} {366}},\ \bibinfo {pages} {102} (\bibinfo {year}
  {2016})}\BibitemShut {NoStop}%
\bibitem [{\citenamefont {Du}\ \emph {et~al.}(2015{\natexlab{a}})\citenamefont
  {Du}, \citenamefont {Bai},\ and\ \citenamefont {Guo}}]{coherenceother7}%
  \BibitemOpen
  \bibfield  {author} {\bibinfo {author} {\bibfnamefont {S.}~\bibnamefont
  {Du}}, \bibinfo {author} {\bibfnamefont {Z.}~\bibnamefont {Bai}}, \ and\
  \bibinfo {author} {\bibfnamefont {Y.}~\bibnamefont {Guo}},\ }\href {\doibase
  10.1103/PhysRevA.91.052120} {\bibfield  {journal} {\bibinfo  {journal} {Phys.
  Rev. A}\ }\textbf {\bibinfo {volume} {91}},\ \bibinfo {pages} {052120}
  (\bibinfo {year} {2015}{\natexlab{a}})}\BibitemShut {NoStop}%
\bibitem [{\citenamefont {Bera}\ \emph {et~al.}(2015)\citenamefont {Bera},
  \citenamefont {Qureshi}, \citenamefont {Siddiqui},\ and\ \citenamefont
  {Pati}}]{coherenceother8}%
  \BibitemOpen
  \bibfield  {author} {\bibinfo {author} {\bibfnamefont {M.~N.}\ \bibnamefont
  {Bera}}, \bibinfo {author} {\bibfnamefont {T.}~\bibnamefont {Qureshi}},
  \bibinfo {author} {\bibfnamefont {M.~A.}\ \bibnamefont {Siddiqui}}, \ and\
  \bibinfo {author} {\bibfnamefont {A.~K.}\ \bibnamefont {Pati}},\ }\href
  {\doibase 10.1103/PhysRevA.92.012118} {\bibfield  {journal} {\bibinfo
  {journal} {Phys. Rev. A}\ }\textbf {\bibinfo {volume} {92}},\ \bibinfo
  {pages} {012118} (\bibinfo {year} {2015})}\BibitemShut {NoStop}%
\bibitem [{\citenamefont {Baumgratz}\ \emph {et~al.}(2014)\citenamefont
  {Baumgratz}, \citenamefont {Cramer},\ and\ \citenamefont
  {Plenio}}]{QuantifyingCoherence}%
  \BibitemOpen
  \bibfield  {author} {\bibinfo {author} {\bibfnamefont {T.}~\bibnamefont
  {Baumgratz}}, \bibinfo {author} {\bibfnamefont {M.}~\bibnamefont {Cramer}}, \
  and\ \bibinfo {author} {\bibfnamefont {M.~B.}\ \bibnamefont {Plenio}},\
  }\href {\doibase 10.1103/PhysRevLett.113.140401} {\bibfield  {journal}
  {\bibinfo  {journal} {Phys. Rev. Lett.}\ }\textbf {\bibinfo {volume} {113}},\
  \bibinfo {pages} {140401} (\bibinfo {year} {2014})}\BibitemShut {NoStop}%
\bibitem [{\citenamefont {Winter}\ and\ \citenamefont
  {Yang}(2016)}]{OperationalResourceTheoryofCoherence}%
  \BibitemOpen
  \bibfield  {author} {\bibinfo {author} {\bibfnamefont {A.}~\bibnamefont
  {Winter}}\ and\ \bibinfo {author} {\bibfnamefont {D.}~\bibnamefont {Yang}},\
  }\href {\doibase 10.1103/PhysRevLett.116.120404} {\bibfield  {journal}
  {\bibinfo  {journal} {Phys. Rev. Lett.}\ }\textbf {\bibinfo {volume} {116}},\
  \bibinfo {pages} {120404} (\bibinfo {year} {2016})}\BibitemShut {NoStop}%
\bibitem [{\citenamefont {Yuan}\ \emph {et~al.}(2015)\citenamefont {Yuan},
  \citenamefont {Zhou}, \citenamefont {Cao},\ and\ \citenamefont
  {Ma}}]{Yuan15intrinsic}%
  \BibitemOpen
  \bibfield  {author} {\bibinfo {author} {\bibfnamefont {X.}~\bibnamefont
  {Yuan}}, \bibinfo {author} {\bibfnamefont {H.}~\bibnamefont {Zhou}}, \bibinfo
  {author} {\bibfnamefont {Z.}~\bibnamefont {Cao}}, \ and\ \bibinfo {author}
  {\bibfnamefont {X.}~\bibnamefont {Ma}},\ }\href {\doibase
  10.1103/PhysRevA.92.022124} {\bibfield  {journal} {\bibinfo  {journal} {Phys.
  Rev. A}\ }\textbf {\bibinfo {volume} {92}},\ \bibinfo {pages} {022124}
  (\bibinfo {year} {2015})}\BibitemShut {NoStop}%
\bibitem [{\citenamefont {Brand\~ao}\ and\ \citenamefont {Gour}(2015)}]{QRT}%
  \BibitemOpen
  \bibfield  {author} {\bibinfo {author} {\bibfnamefont {F.~G. S.~L.}\
  \bibnamefont {Brand\~ao}}\ and\ \bibinfo {author} {\bibfnamefont
  {G.}~\bibnamefont {Gour}},\ }\href {\doibase 10.1103/PhysRevLett.115.070503}
  {\bibfield  {journal} {\bibinfo  {journal} {Phys. Rev. Lett.}\ }\textbf
  {\bibinfo {volume} {115}},\ \bibinfo {pages} {070503} (\bibinfo {year}
  {2015})}\BibitemShut {NoStop}%
\bibitem [{\citenamefont {Streltsov}\ \emph {et~al.}(2015)\citenamefont
  {Streltsov}, \citenamefont {Singh}, \citenamefont {Dhar}, \citenamefont
  {Bera},\ and\ \citenamefont
  {Adesso}}]{MeasuringQuantumCoherencewithEntanglement}%
  \BibitemOpen
  \bibfield  {author} {\bibinfo {author} {\bibfnamefont {A.}~\bibnamefont
  {Streltsov}}, \bibinfo {author} {\bibfnamefont {U.}~\bibnamefont {Singh}},
  \bibinfo {author} {\bibfnamefont {H.~S.}\ \bibnamefont {Dhar}}, \bibinfo
  {author} {\bibfnamefont {M.~N.}\ \bibnamefont {Bera}}, \ and\ \bibinfo
  {author} {\bibfnamefont {G.}~\bibnamefont {Adesso}},\ }\href {\doibase
  10.1103/PhysRevLett.115.020403} {\bibfield  {journal} {\bibinfo  {journal}
  {Phys. Rev. Lett.}\ }\textbf {\bibinfo {volume} {115}},\ \bibinfo {pages}
  {020403} (\bibinfo {year} {2015})}\BibitemShut {NoStop}%
\bibitem [{\citenamefont {Girolami}(2014)}]{Giro14observable}%
  \BibitemOpen
  \bibfield  {author} {\bibinfo {author} {\bibfnamefont {D.}~\bibnamefont
  {Girolami}},\ }\href {\doibase 10.1103/PhysRevLett.113.170401} {\bibfield
  {journal} {\bibinfo  {journal} {Phys. Rev. Lett.}\ }\textbf {\bibinfo
  {volume} {113}},\ \bibinfo {pages} {170401} (\bibinfo {year}
  {2014})}\BibitemShut {NoStop}%
\bibitem [{\citenamefont {Du}\ \emph {et~al.}(2015{\natexlab{b}})\citenamefont
  {Du}, \citenamefont {Bai},\ and\ \citenamefont {Qi}}]{Du15coherence}%
  \BibitemOpen
  \bibfield  {author} {\bibinfo {author} {\bibfnamefont {S.}~\bibnamefont
  {Du}}, \bibinfo {author} {\bibfnamefont {Z.}~\bibnamefont {Bai}}, \ and\
  \bibinfo {author} {\bibfnamefont {X.}~\bibnamefont {Qi}},\ }\href
  {http://dl.acm.org/citation.cfm?id=2871378.2871381} {\bibfield  {journal}
  {\bibinfo  {journal} {Quantum Info. Comput.}\ }\textbf {\bibinfo {volume}
  {15}},\ \bibinfo {pages} {1307} (\bibinfo {year}
  {2015}{\natexlab{b}})}\BibitemShut {NoStop}%
\bibitem [{\citenamefont {Ma}\ \emph {et~al.}(2016)\citenamefont {Ma},
  \citenamefont {Yadin}, \citenamefont {Girolami}, \citenamefont {Vedral},\
  and\ \citenamefont {Gu}}]{Ma16converting}%
  \BibitemOpen
  \bibfield  {author} {\bibinfo {author} {\bibfnamefont {J.}~\bibnamefont
  {Ma}}, \bibinfo {author} {\bibfnamefont {B.}~\bibnamefont {Yadin}}, \bibinfo
  {author} {\bibfnamefont {D.}~\bibnamefont {Girolami}}, \bibinfo {author}
  {\bibfnamefont {V.}~\bibnamefont {Vedral}}, \ and\ \bibinfo {author}
  {\bibfnamefont {M.}~\bibnamefont {Gu}},\ }\href {\doibase
  10.1103/PhysRevLett.116.160407} {\bibfield  {journal} {\bibinfo  {journal}
  {Phys. Rev. Lett.}\ }\textbf {\bibinfo {volume} {116}},\ \bibinfo {pages}
  {160407} (\bibinfo {year} {2016})}\BibitemShut {NoStop}%
\bibitem [{\citenamefont {Chitambar}\ \emph {et~al.}(2016)\citenamefont
  {Chitambar}, \citenamefont {Streltsov}, \citenamefont {Rana}, \citenamefont
  {Bera}, \citenamefont {Adesso},\ and\ \citenamefont
  {Lewenstein}}]{AssistedDistillationofQuantumCoherence}%
  \BibitemOpen
  \bibfield  {author} {\bibinfo {author} {\bibfnamefont {E.}~\bibnamefont
  {Chitambar}}, \bibinfo {author} {\bibfnamefont {A.}~\bibnamefont
  {Streltsov}}, \bibinfo {author} {\bibfnamefont {S.}~\bibnamefont {Rana}},
  \bibinfo {author} {\bibfnamefont {M.~N.}\ \bibnamefont {Bera}}, \bibinfo
  {author} {\bibfnamefont {G.}~\bibnamefont {Adesso}}, \ and\ \bibinfo {author}
  {\bibfnamefont {M.}~\bibnamefont {Lewenstein}},\ }\href {\doibase
  10.1103/PhysRevLett.116.070402} {\bibfield  {journal} {\bibinfo  {journal}
  {Phys. Rev. Lett.}\ }\textbf {\bibinfo {volume} {116}},\ \bibinfo {pages}
  {070402} (\bibinfo {year} {2016})}\BibitemShut {NoStop}%
\bibitem [{\citenamefont {Qi}\ \emph {et~al.}(2013)\citenamefont {Qi},
  \citenamefont {Hou}, \citenamefont {Li}, \citenamefont {Dong}, \citenamefont
  {Xiang},\ and\ \citenamefont {Guo}}]{Qi13quantum}%
  \BibitemOpen
  \bibfield  {author} {\bibinfo {author} {\bibfnamefont {B.}~\bibnamefont
  {Qi}}, \bibinfo {author} {\bibfnamefont {Z.}~\bibnamefont {Hou}}, \bibinfo
  {author} {\bibfnamefont {L.}~\bibnamefont {Li}}, \bibinfo {author}
  {\bibfnamefont {D.}~\bibnamefont {Dong}}, \bibinfo {author} {\bibfnamefont
  {G.}~\bibnamefont {Xiang}}, \ and\ \bibinfo {author} {\bibfnamefont
  {G.}~\bibnamefont {Guo}},\ }\href@noop {} {\bibfield  {journal} {\bibinfo
  {journal} {Sci. Rep.}\ }\textbf {\bibinfo {volume} {3}},\ \bibinfo {pages}
  {3496} (\bibinfo {year} {2013})}\BibitemShut {NoStop}%
\bibitem [{\citenamefont {Miranowicz}(2004)}]{Miranowicz2004272}%
  \BibitemOpen
  \bibfield  {author} {\bibinfo {author} {\bibfnamefont {A.}~\bibnamefont
  {Miranowicz}},\ }\href {\doibase
  http://dx.doi.org/10.1016/j.physleta.2004.05.001} {\bibfield  {journal}
  {\bibinfo  {journal} {Physics Letters A}\ }\textbf {\bibinfo {volume}
  {327}},\ \bibinfo {pages} {272 } (\bibinfo {year} {2004})}\BibitemShut
  {NoStop}%
\bibitem [{\citenamefont {Kwiat}\ \emph {et~al.}(1999)\citenamefont {Kwiat},
  \citenamefont {Waks}, \citenamefont {White}, \citenamefont {Appelbaum},\ and\
  \citenamefont {Eberhard}}]{Kwia99Ultrabright}%
  \BibitemOpen
  \bibfield  {author} {\bibinfo {author} {\bibfnamefont {P.~G.}\ \bibnamefont
  {Kwiat}}, \bibinfo {author} {\bibfnamefont {E.}~\bibnamefont {Waks}},
  \bibinfo {author} {\bibfnamefont {A.~G.}\ \bibnamefont {White}}, \bibinfo
  {author} {\bibfnamefont {I.}~\bibnamefont {Appelbaum}}, \ and\ \bibinfo
  {author} {\bibfnamefont {P.~H.}\ \bibnamefont {Eberhard}},\ }\href@noop {}
  {\bibfield  {journal} {\bibinfo  {journal} {Phys. Rev. A}\ }\textbf {\bibinfo
  {volume} {60}},\ \bibinfo {pages} {R773} (\bibinfo {year}
  {1999})}\BibitemShut {NoStop}%
\end{thebibliography}%

{\bf Acknowledgements}

We thank Jayne Thompson, Mile Gu and Vlatko Vedral for fruitful discussions. This work was supported by National Natural Science Foundation of China (Grant No. 11574291, 61108009, 61222504) and National Key R \& D Program (Grant No. 2016YFA0301700).

{\bf Author contributions}

G.-Y.X. devised the experiment. G.-Y.X., Z.H. designed the experiment. Z.H and K.-D.W implement the experiment assisted by H.-S.Z and Y.Y, K.-D.W collected and analyzed the experimental data and wrote the manuscript with the help of Z.H., G.-Y.X., C.-F.L and G.-C.G.. All authors contribute to the revision of the manuscript. G.-Y.X. supervises the project.

{\bf Additional information}

\newpage
\section{\label{sec:supplemental material}supplemental material}

\subsection{\label{sec:append:Selection of measurement basis}Selection of Measurement Basis}
In the following, we will show how to select optimal measurement basis when the shared state is a pure two-qubit state $\rho^{AB}$.

According to theoretical analysis \cite{AssistedDistillationofQuantumCoherence}, we expand the purification $\ket{\Psi}^{AB}$ in the incoherent basis
\begin{equation}\label{expandABpure}
\ket{\Psi}^{AB}=\sum^{1}_{k=1}c_{k}\ket{\Psi_k}^{A}\ket{k}^{B},
\end{equation}
then there always exist orthogonal states $|\eta\pm\rangle$ which are mutually unbiased with respect to the two state $|\Psi_{k}\rangle^{A}$. When Alice performs a von Neumann measurement in the $|\eta\pm\rangle$ basis, Bob will be in one of two post-measurement state $|\Phi_{\pm}\rangle^{B}$ associated with the outcome of the measurement on Alice, in both cases, the state has coherence $C_{r}(|\Phi_{\pm}\rangle^{B})=S(\Delta(\rho^{B}))$ and with $C^{\infty}_{a}(\rho)=S(\Delta(\rho))$, we can find \cite{AssistedDistillationofQuantumCoherence}
\begin{equation}\label{singlecopy}
C_{a}(\rho)=C^{\infty}_{a}(\rho)=S(\Delta(\rho)),
\end{equation}
for pure states the coherence of assistance is equal to the coherence of collaboration in the asymptotic setting, $C^{A|B}_{d}(\ket{\Psi}^{AB})=C^{\infty}_{a}(\rho^{B})$. Moreover, when Bob¡¯s system is a qubit and the overall state is pure, the coherence of assistance and the coherence of collaboration are equivalent even in the single-copy case, $C_{a}(\rho)=C^{\infty}_{a}(\rho)$.

Hence we can take use of the theoretical analysis to maximize the distillable coherence on Bob's system. In the case where the shared state is a pure two-qubit state, when we perform von Neumann measurement on Alice in the mutually unbiased basis, we can obtain the maximal increase in distillable coherence in Bob's system. And in our experiments, when we expand the two pure states in incoherent basis, we obtain the sets of states $|\Psi_{k}\rangle^{A}$ are $\ket{H}$ , $\ket{V}$ for state $\ket{\Psi}^{AB}=\cos2\theta\ket{HH}+\sin2\theta\ket{VV}$ and $\cos2\theta\ket{H}\pm\sin2\theta\ket{V}$ for state $\ket{\Psi}^{AB}=\frac{1}{\sqrt{2}}(\cos2\theta\ket{HH}+\cos2\theta\ket{HV}+\sin2\theta\ket{VH}-\sin2\theta\ket{VV})$. And obviously there exists a shared mutually unbiased basis $\ket{y\pm}$ with respect to both cases. So we can apply von Neumann measurement on Alice along $|y\pm\rangle$ basis in the first part of our experiments.

 In the experiments for Werner states, we considered the task in one-copy scenario because the complex collective measurement on many copies of a state is hard to be implemented in the presence of experimental limitation. As the two qubit Werner states have a form of $\rho^{AB}=p\ketbra{\Psi^-}{\Psi^-}+(1-p)\frac{I}{4}$ with high symmetry, thus when Alice measures her system in the basis $\ket{\eta_\pm}=\frac{1}{\sqrt{2}}(\ket{H}\pm e^{i\varphi}\ket{V})$ with arbitrary $\varphi$, Bob's state will collapse to $\rho^{B}_{1}=p\ketbra{\eta_\mp}{\eta_\mp}+(1-p)\frac{I}{2}$, and we chose the same $\ket{y_\pm}$ as that for pure states.

\subsection{\label{sec:append:Experimental apparatus}Experimental Apparatus}
We now move to the detailed experimental arrangement in our laboratory based on the polarization-entangled photon pairs.

In state preparation module in Fig.~\ref{experimental_setup}, two type I phase-matched $\beta$-barium borate (BBO) crystals, whose optic axes are normal to each other \cite{Kwia99Ultrabright}, are pumped by the continuous $\textmd{Ar}^{+}$ laser at 351.1 nm with power up to 60 mW for the generation of photon pairs with a central wavelength at $\lambda$=702.2 nm via spontaneous parametric down conversion process (SPDC). HWP1 working at $\lambda$=351.1 nm is set in front of the BBO crystals to control the quantum state of the generated photon pairs encoded in polarization. Half-wave plates at both ends of the two single mode fibers (SMF) are used to control polarization. In one branch, QWP3 is tilted to compensate the phase of the two-photon state for the generation of required states. This setup is capable of preparing two classes of pure states $\ket{\Psi}^{AB}=\cos2\theta\ket{HH}+\sin2\theta\ket{VV}$ and $\ket{\Psi}^{AB}=\frac{1}{\sqrt{2}}(\cos2\theta\ket{HH}+\cos2\theta\ket{HV}+\sin2\theta\ket{VH}-\sin2\theta\ket{VV})$, where $\theta$ is the rotation angle of HWP1.

As for the generation of Werner states, two 50/50 beam splitters (BS) are inserted into one branch. In the transmission path, the two-photon state is prepared as the singlet state $\ket{\Psi^-}=\frac{1}{\sqrt{2}}(\ket{HV}-\ket{VH})$ when the rotation angle of HWP1 is set as $\theta=22.5^\circ$. In the reflected path, three 446$\lambda$ quartz crystals and a half wave plate with $22.5^{\circ}$ are used to dephase the two-photon state into a completely mixed state $\frac{I}{4}$. The ratio of the two states mixed at the output port of the second BS can be changed by the two adjustable apertures for the generation of arbitrary Werner states with $\rho^{AB}=p\ketbra{\Psi^-}{\Psi^-}+(1-p)\frac{I}{4}$. Out of the state preparation module, the two photons are distributed to Alice and Bob as shown in Fig.~\ref{experimental_setup}.

In the LQICC module, on Alice's side, QWP1, HWP3, PBS and two single photon detectors are used to perform arbitrary assisted projective measurements and the measurement results are sent to Bob via classical communication channel. In the next stage, QWP2, HWP4, PBS and two single photon detectors are used to perform tomography on Bob's photon. Actually, we used a beam displacer (BD) together with a $45^{\circ}$ holophote which can act as a PBS and has a rather high extinction ratio.

\subsection{\label{sec:append:Experimental Data}Experimental Data}
\begin{table*}[!htbp]
\caption{Experimental data for pure states $\ket{\Psi}^{AB}=\cos2\theta\ket{HH}+\sin2\theta\ket{VV}$}
\begin{tabular}{c|cccccccccc}
\hline
\hline
$\theta(^{\circ})$&0.0&2.5&5.0&7.5&10.0&12.5&15.&17.5&20.0&22.5\\
\hline
$C_{d}(\rho^{B})$&0.0335&0.0193&0.0326&0.0136&0.00161&0.00464&0.010953&0.00324&0.00222&0.00373\\
\hline
$C_{d}^{A|B}(\ket{\Psi}^{AB})$&0.0394&0.105&0.179&0.319&0.553&0.694&0.778&0.891&0.927&0.905\\
\hline
$\delta C_{d}(\rho^{B})$&0.00587&0.0854&0.146&0.306&0.552&0.690&0.767&0.888&0.925&0.902\\
\hline
\hline
 $\theta(^{\circ})$&25.0&27.5&30.0&32.5&35.0&37.5&40.&42.5&45.0& \\
\hline
 $C_{d}(\rho^{B})$&0.00794&0.00760&0.00754&0.00919&0.0171&0.0153&0.0208&0.0153&0.0432&\\
\hline
 $C_{d}^{A|B}(\ket{\Psi}^{AB})$&0.885&0.765&0.707&0.559&0.384&0.276&0.177&0.0776&0.0599&\\
\hline
 $\delta C_{d}(\rho^{B})$&0.877&0.757&0.699&0.550&0.367 &0.261&0.156&0.0623&0.0167\\
\hline
\hline
\end{tabular}
\end{table*}

\begin{table*}[!htbp]
\caption{Experimental data for pure states $\ket{\Psi}^{AB}=\frac{1}{\sqrt{2}}(\cos2\theta\ket{HH}+\cos2\theta\ket{HV}+\sin2\theta\ket{VH}-\sin2\theta\ket{VV})$}
\begin{tabular}{c|cccccccccc}
\hline
\hline
$\theta(^{\circ})$&0.0&2.5&5.0&7.5&10.0&12.5&15.0&17.5&20.0&22.5\\
\hline
$C_{d}(\rho^{B})$&0.949&0.886&0.802&0.633&0.462&0.297&0.141&0.0698&0.0166&0.00606\\
\hline
$C_{d}^{A|B}(\ket{\Psi}^{AB})$&0.915&0.916&0.948&0.962&0.950&0.868&0.875&0.885&0.860&0.914\\
\hline
$\delta C_{d}(\rho^{B})$&-0.0342&0.0299&0.146&0.329&0.488&0.571&0.734&0.815&0.843&0.908\\
\hline
\hline
 $\theta(^{\circ})$&25.0&27.5&30.0&32.5&35.0&37.5&40.0&42.5&45.0& \\
\hline
 $C_{d}(\rho^{B})$&0.0334&0.0879&0.239&0.343&0.518 &0.647&0.813&0.938&0.944& \\
\hline
 $C_{d}^{A|B}(\ket{\Psi}^{AB})$&0.886&0.884&0.929&0.889&0.910&0.889&0.956&0.919&0.912& \\
\hline
 $\delta C_{d}(\rho^{B})$&0.853&0.796&0.690&0.546&0.393&0.242&0.143&-0.0190&-0.0325&\\
\hline
\hline
\end{tabular}
\end{table*}

\begin{table*}[!htbp]
\caption{Experimental data for Werner states}
\begin{tabular}{c|cccccccc}
\hline
\hline
$P$&0.100&0.164&0.280&0.363&0.397&0.429&0.516&0.541\\
\hline
$C_{d}(\rho^{B}_{0})$&0.00503&0.00349&0.00826&0.00676&0.00889&0.00644&0.00442&0.00471\\
\hline
$C_{d}(\rho^{B}_{1})$&0.0122&0.0155&0.0659&0.0830&0.106&0.128&0.212&0.230\\
\hline
$\delta C_{d}(\rho^{B})$&0.00716&0.0120&0.0577&0.0762&0.0968&0.121&0.208&0.225\\
\hline
\hline
$P$&0.598&0.630&0.689&0.751&0.808&0.878&0.895&0.949\\
\hline
$C_{d}(\rho^{B}_{0})$&0.00536&0.00391&0.00468&0.00996&0.000813&0.00554&0.00313&0.00418\\
\hline
$C_{d}(\rho^{B}_{1})$&0.291&0.308&0.324&0.528&0.521&0.672&0.735&0.762\\
\hline
$\delta C_{d}(\rho^{B})$&0.286&0.304&0.320&0.518&0.520&0.667&0.732&0.758\\
\hline
\hline
\end{tabular}
\end{table*}
In this section we show the detail experimental data.




\end{document}